\documentclass[aps,amssymb,amsmath,amsbsy,amsfonts,pra,showpacs]{article}
\usepackage[utf8]{inputenc}
\usepackage{amsthm}
\usepackage{color, graphicx, enumerate}
\usepackage{authblk}
\usepackage{amsmath}
\usepackage{bbm}

\newtheorem*{thm*}{Theorem}
\newtheorem*{lm*}{Lemma}
\newtheorem*{claim*}{Claim}

\newtheorem{proposition}{Proposition}
\newtheorem{property}{Property}
\newtheorem{definition}{Definition}

\theoremstyle{definition}

\newcommand{\e}{\varepsilon}

\newcommand{\1}{\textbf{1}}

\begin{document}
\title{Selfcomplementary quantum channels}

\author{Marek Smaczy{\'n}ski$^1$, Wojciech Roga$^2$, Karol {\.Z}yczkowski$^{1,3}$}
\affil{$^1$Smoluchowski Institute of Physics, Jagiellonian University, \L{}ojasiewicza 11, 30-348 Cracow, Poland. \texttt{marek.smaczynski@uj.edu.pl}\\
$^2$Department of Physics, University of Strathclyde, John Anderson Building, 107 Rottenrow, Glasgow G4 0NG, UK. \texttt{wojciech.roga@strath.ac.uk}\\
$^3$Center of Theoretical Physics, Polish Academy of Sciences, Al. Lotnikow 32/46, 02-668 Warsaw, Poland. \texttt{karol@cft.edu.pl}}

\maketitle


\begin{abstract}
Selfcomplementary quantum channels are characterized by such an interaction between 
the principal quantum system and the environment that leads to
the same output states of both interacting systems.
These maps can describe approximate quantum copy machines, 
as  perfect copying of an unknown quantum state is not possible 
due to the celebrated no--cloning theorem.
We provide here a parameterization of a large class of selfcomplementary
channels and analyze their properties.
Selfcomplementary channels preserve some residual coherences and residual entanglement.
Investigating some measures of non-Markovianity we show that time
evolution under selfcomplementary channels is highly non-Markovian.
\end{abstract}

\section{Introduction}\label{sec:introduction}

Capacity of a noisy information channel characterizes the amount of information per one symbol 
which is reliably transmitted through the channel
in the limit of a long message send \cite{Holevo2012}.
This general definition can be made more precise,
 as one specifies what kind of information 
is transmitted and which additional resources can be used.
For a discussion on different classes of channel capacities
see e.g.~\cite{Bennett2002}.

In particular, the capacity of a classical channel characterizes the average number of classical bits of information that can be reliably transmitted through the channel 
in a long sequence of symbols. Alternatively, it refers to the average 
dimensionality of a discrete vector space 
such that every vector of symbols from this space  
transmitted through the channel
can be recovered with a high fidelity with a help of a suitable error correction scheme.
Analogously, the quantum capacity $Q$ of a quantum channel characterizes 
the average number of
 qubits per a single use of the channel that can be reliably recovered 
from long sequences of transmitted states. 
Alternatively, this capacity characterizes the average dimensionality 
of the Hilbert subspace 
such that every quantum state belonging to 
this subspace can be transmitted through the channel and recovered with a vanishing error.
In consequence, a quantum channel of a positive quantum  capacity can 
preserve  coherent superposition of states or quantum entanglement at least 
for some quantum states.

The action of a quantum channel can be modeled by an interaction of a quantum system with an environment.
Capacity of a quantum channel can be expressed~\cite{Devetak2005}
in terms of the {\sl coherent information}, 
defined as the difference between the von Neumann entropy of the output state 
and the entropy of the environment after the evolution~\cite{Schumacher1996}
-- see
 Sec.~\ref{definitions}. 
A transformation which maps an input state into the state of the environment after the evolution is called the {\sl complementary channel}~\cite{Holevo2012,Holevo,Petz, Wilde}.

Although the definitions of the classical and the quantum channel capacities are 
similar, these two notions differ in several ways.
To show this consider the {\sl dephasing channel}, which for a given basis
removes all off--diagonal elements of the density matrix.
This channel transforms any coherent superposition of pure orthogonal states into their
 statistical mixture, however, any classical state
remains unchanged.
Therefore, the classical capacity of this channel can be positive, 
while its quantum capacity is equal to zero, as there does not exist even a two-dimensional Hilbert subspace which survives the action of the channel~\cite{Lloyd}.

In the present work we study a family of {\sl selfcomplementary} quantum
channels, which transform an input state and an initial state 
of the environment into two identical states. 
By definition, the coherent information of such a channel and its quantum capacity 
are equal to zero, while its classical capacity can be positive.
The class of selfcomplementary channels contains, for instance,
the dephasing channel.
We show that in contrast to the dephasing channel,
a generic selfcomplementary channel 
is not entanglement breaking~\cite{Horodecki2003},
as it can preserve some residual coherences. 
The fact that the quantum channel capacity of a selfcomplementary channel
is equal to zero can be related with 
the famous no--cloning theorem, see Sec.~\ref{sec:remarks}.
Since the no--cloning theorem does not hold for classical states,
which are orthogonal or coincide, 
the classical channel capacity of a selfcomplementary channel
can be positive.

We study also memory effects induced by the time evolution
under the action of selfcomplementary channels.
Investigations of non-Markovian quantum evolutions 
and various measures of non-Markovianity 
attracted recently a lot of attention
~\cite{Rivas2010,Chruscinski2012,Bylicka2014,Rivas2014,Addis2014,Torre2015}. 
Memory effects of quantum evolutions
may increase efficiency 
of some of the quantum protocols~\cite{Laine2014}
or influence the time evolution of 
 biological systems~\cite{Huelga2013}. 
Selfcomplementary channels provide examples of highly non-Markovian evolution,
and this property can be detected investigating  
the residual entanglement~\cite{Rivas2010}.

The paper is organized as follows. 
In Sec.~\ref{definitions} we review basic definitions related to 
quantum channels and their capacities.
Selfcomplementary channels and their key properties are discussed in Sec.~\ref{selfie}. In particular, we show lower and upper
bounds for the entropy of selfcomplementary maps.
A parameterization of the set of one-qubit selfcomplementary channels is given in Sec.~\ref{sectionSCN2} and is
generalized for higher dimensions in Appendices~\ref{gutritgen} and~\ref{sectionSCNall}. 
Residual entanglement is analyzed in Sec.~\ref{resent}, while 
relations between the no--cloning theorem and the zero quantum capacity 
of selfcomplementary channels is discussed in Sec.~\ref{sec:remarks}. 
Proofs of propositions formulated in the main body of the paper
 are relegated to Appendices.

\section{Quantum channels, coherent information and channel capacity}\label{definitions}

Time evolution of an open quantum system $\cal S$ can be described in terms of a global unitary dynamics $U$, which couples the quantum system with an environment ${\cal E}$
~\cite{Stinespring1955}.
Performing  partial trace over the environment 
one defines a linear quantum map $\Phi$, which acts
on the principal system, 
%
\begin{equation}
\rho'=\Phi(\rho)={\rm Tr}_{\cal E}[U(\rho \otimes \sigma)U^{\dagger}],
\label{EnvirnomentRepresentation}
\end{equation}
where $\sigma$ denotes an initial state of the environment ${\cal E}$.
%
Any evolution $\Phi$ of the above form 
preserves positivity of the input state.
Furthermore, $\Phi$ belongs to the class of {\sl completely positive} (CP) maps,
as its extension on an arbitrary larger space, $\Phi\otimes {\mathbbm I}$, preserves positivity.
Any CP map $\Phi$ which preserves normalization of the state 
is called 
a {\sl stochastic map}, quantum channel or quantum operation.
It is well known~\cite{Stinespring1955} that any stochastic map 
admits a unitary representation~(\ref{EnvirnomentRepresentation}).

It is legitimate to ask about a fate of the environment
after the interaction with the principal system.
The corresponding evolution of the state of the environment reads
\begin{equation}
\sigma'=\widetilde{\Phi}(\rho)={\rm Tr}_{\cal S}[U(\rho \otimes \sigma)U^{\dagger}].
\end{equation}
The map $\widetilde{\Phi}$ defined in this way forms channel 
{\sl complementary} to $\Phi$.

To characterize information which can be encoded in a quantum state $\rho$ 
one often uses its von Neumann entropy, $S(\rho)=- {\rm Tr} ( \rho \log \rho)$.
This quantity can also be applied to  describe properties of quantum channels. 
The {\sl coherent information} which is transmitted through a channel $\Phi$
acting on the initial state $\rho$ is defined~\cite{Schumacher1996} as 
\begin{equation}
I_{coh}(\Phi,\rho)= S\bigl(\Phi(\rho)\bigr)-S\bigl(\tilde{\Phi}(\rho)\bigr) .
\label{kohinf}
\end{equation}
For classical states, the coherent information 
takes negative values only. 
However,
if $\rho$ is a quantum state, the coherent information $I_{coh}$ can also be positive,  
so it can be used to quantify, how well the quantum coherences are preserved by the channel~\cite{Haiashi2006}. 
Coherent information is monotonically decreasing with respect to a concatenation of the channels and this property 
is often referred to as the data processing inequality. Furthermore, it  is convex with respect to linear combinations of the channels and concave with respect to linear combinations of the states -- see~\cite{Haiashi2006} and references therein. 
Moreover,  
coherent information maximized over the input states is not additive with respect to tensor product of two channels~\cite{Divincenzo1998}

\begin{equation}
\max_{\rho^{AB}}I_{coh}(\Phi^A\otimes\Phi^B,\rho^{AB})\geq \max_{\rho^{A}}I_{coh}(\Phi^A,\rho^{A})+\max_{\rho^B}I_{coh}(\Phi^B,\rho^{B}).
\label{additive}
\end{equation}

For any quantum channel $\Phi$ one defines its {\sl quantum capacity} 
\begin{equation}
Q_C\equiv \lim_{n\rightarrow\infty}\sup\frac{\log{d}}{n}, 
  \label{channelcap1}
\end{equation}
where $d$ and $n$ are such that there exists a $d$--dimensional subspace ${\mathcal S}\subseteq {\mathcal H}_{input}^{\otimes n}$ and there exist such coding and error correcting schemes 
that every input state from ${\mathcal S}$ is transmitted through the $n$ copies of the channel with arbitrary high fidelity. 
The definition of the capacity requires to analyze the coding and decoding schemes in Hilbert spaces of asymptotically large dimensions. 
However, the capacity can be related with coherent information of the channel $\Phi$ used in parallel $n$ times~\cite{Devetak2005}
%
\begin{equation}
  Q_C= \lim_{n \rightarrow \infty} \max_{\rho} \frac{1}{n} I_{coh}(\Phi^{\otimes n},\rho).
  \label{channelcap}
\end{equation}  

In the subsequent section, we will analyze a class of quantum channels for which one shot coherent information is zero and we will discuss the corresponding quantum channel capacity.

\section{Selfcomplementary channels, definition and properties}\label{selfie}

Let us define a class of selfcomplementary channels:

\noindent
\begin{definition} \label{SCdefinition}
A quantum channel $\Phi_{self}$ is called a selfcomplementary channel if for every input state an output of the channel is identical with an output of its complementary counterpart, i.e., \begin{equation}
\Phi_{self}=\widetilde{\Phi}_{self}
\label{GenSym}
\end{equation}
for properly chosen bases of the two output states.
\end{definition}
These channels have been studied earlier in~\cite{Smith2008}, where they are called symmetric side channels. 
Before characterizing selfcomplementary channels in detail, let us discuss a relation between Kraus operators (see Appendix~\ref{trombalski}) associated with a quantum channel and its complementary counterpart given in the following Proposition. 

\begin{proposition}\label{propkraus} Denote a set of density matrices of an $n$ level system as ${\mathcal M}_n$. Assume that a quantum channel $\Phi: {\mathcal M}_N\rightarrow {\mathcal M}_M$ is represented by Kraus operators $K^i$ as follows, $\Phi(\rho)=\sum_{i=1}^{k}K^i\rho K^{i\dagger}$ and the Kraus representation of the complementary channel $\widetilde{\Phi}(\rho):{\mathcal M}_N\rightarrow {\mathcal M}_k$ is given by $\widetilde{\Phi}(\rho)=\sum_{i=1}^{M}\tilde{K}^i\rho \tilde{K}^{i\dagger}$, where the dimensionality of the input system, the output system and the environment are $N$, $M$ and $k$ respectively. The following relation between the Kraus operators associated to these channels holds true
\begin{equation}
\widetilde{K}^{\alpha}_{ij}=K^{i}_{\alpha j},\qquad i=1,...,k,\qquad \alpha=1,...,M,\qquad j=1,...,N,
\label{pantalon}
\end{equation}
where the lower indexes indicate the matrix entries and the upper indices numerate the Kraus operators.
\end{proposition}
The proof is given in Appendix~\ref{appkraus}. Proposition~\ref{propkraus} implies that if a given quantum channel $\Phi$ is defined by $k$ Kraus operators represented by $M\times N$ matrices
$$\Phi:\{K_{i=1}^k\} \rightarrow M\overbrace{
\underbrace{\begin{cases}\begin{bmatrix}&&\\&...&\\&&\end{bmatrix}\end{cases}}_N,
\begin{bmatrix}&&\\&...&\\&&\end{bmatrix},...,
\begin{bmatrix}&&\\&...&\\&&\end{bmatrix}}^k,$$ 
the complementary channel $\widetilde{\Phi}$ is characterized by $M$ Kraus operators given by $k \times N$ matrices
$$\widetilde{\Phi}:\{\widetilde{K}_{i=1}^N\} \rightarrow k\overbrace{
\underbrace{\begin{cases}\begin{bmatrix}&&\\&...&\\&&\end{bmatrix}\end{cases}}_N,...,
\begin{bmatrix}&&\\&...&\\&&\end{bmatrix}}^M.$$

In consequence, in order to satisfy the equality between $\Phi_{self}$ and  $\widetilde{\Phi}_{self}$ for a selfcomplementary map, the dimensionality of the environment has to be equal to the dimensionality of the input state, i.e., $N=k$. This necessary condition for selfcomplementarity of a channel can be expressed also in terms of the so-called Choi-Jamio{\l}kowski state corresponding to the channel 
\begin{equation}
\frac{1}{N}D_{\Phi}=[\1_N\otimes\Phi]\big(|\phi^+\left.\right\rangle\left\langle\right. \phi^+|\big),
\label{choijam}
\end{equation}
where
$|\phi^+\left.\right\rangle=\frac{1}{\sqrt{N}}\sum_{i=1}^N |i\left.\right\rangle\otimes|i\left.\right\rangle$ is a maximally entangled state. The rank of the Choi-Jamio{\l}kowski state is called the rank of the channel and it determines 
the smallest number of the Kraus operators necessary to represent the map.
In general the rank $R$ of the channel satisfies relations $1 \le R \le N^2$.
However, for a selfcomplementary channel $\Phi_{self}:{\mathcal M}_N\rightarrow {\mathcal M}_N$ we have 
\begin{equation}
{\rm Rank}(D_{\Phi_{self}})=N.
\label{PropertyRankN}
\end{equation}
The following examples show some consequences of this statement. A single-qubit depolarizing map is defined as a channel that projects any state into a maximally mixed state. This channel has zero quantum capacity, but it does not belong to the class of selfcomplementary channels because its rank is 4, whereas the dimensionality of the input state is 2. Similarly, the identity channel or any single-qubit unitary channel that has rank one cannot be selfcomplementary. In consequence, selfcomplementary channels cannot be neither very noisy nor reversible.

The definition~\ref{SCdefinition} and the definition of coherent information given in Eq.~(\ref{kohinf}) imply that
 $I_{coh}(\Phi_{self},\rho)$ is equal to zero for any initial state $\rho$. It does not guarantee, however, that the quantum capacity Eq.~(\ref{channelcap}) is also $0$, since the coherent information is not additive, see Eq.~(\ref{additive}). On the other hand, the zero quantum capacity of these channels is justified by the following Proposition proved in Appendix~\ref{appproduct}.
\begin{proposition} \label{PropentTensor}
The tensor product of two selfcomplementary channels is also selfcomplementary,
\begin{equation}
  \Phi_{self} \otimes \Psi_{self} = \Lambda_{self}.
\end{equation}
\end{proposition}
Therefore, the quantum capacity is additive with respect to the tensor product and equal to $0$ for all selfcomplementary channels. Eventually, let us also emphasize the following property
\begin{property}\label{PropConc}
Concatenation of two arbitrary selfcomplementary channels does not need to be selfcomplementary.
However, the quantum channel capacity of any composition of these channels is equal to zero.
\end{property}
The statement is justified as follows. The number  of the Kraus operators in a composition of two selfcomplementary channels is different than the number of the Kraus operators corresponding to one of them. Therefore, the dimensionality of the environment needed to represent this composition is greater than the dimensionality of an input state. Due to Eq.~(\ref{PropertyRankN}) the concatenation is in general not selfcomplementary anymore. The second statement of Property~\ref{PropConc} is derived from the data processing inequality~\cite{Schumacher1996}. It states that a composition of two channels cannot increase the coherent information above the value related with the first of these channels. This implies  the zero quantum channel capacity for any concatenation of selfcomplementary channels.

Although, zero quantum capacity implies that there is no subspace that can be exactly transmitted through the channel, it does not mean that these channels completely destroy coherences or even entanglement. Indeed, the coherences are diminished, but they do not vanish entirely.
Therefore, in the following sections we study the 
impact of the selfcomplementary channels on quantum coherences and quantum entanglement.

\subsection{Decohering properties of selfcomplementary channels and their classical channel capacity}\label{secentropy}

The entropy $S^{map}$ of a quantum channel $\Phi$ is  defined  \cite{Roga2011,Roga2013}
as the von Neumann entropy of the corresponding
Choi-Jamio{\l}kowski state  $D_{\Phi}/N$ given in Eq.~(\ref{choijam}). 
 Since the vanishing map entropy characterizes reversible unitary channels, while its maximum 
is achieved for maximally depolarizing channels, this quantity 
describes the degree of decoherence induced by the particular quantum channel. 
 Selfcomplementarity of a channel implies the following properties on its entropy $S(\Phi)$.
\begin{proposition} \label{propent}
a) The map entropy of a selfcomplementary channel $\Phi_{self}: {\mathcal M}_N\rightarrow {\mathcal M}_N$ is equal to the entropy of an image of the maximally mixed state, i.e.,
\begin{equation}
 S^{map}(\Phi_{self})=S\left(\Phi_{self}(\rho_*)\right).
\end{equation}
b) The map entropy of the selfcomplementary channel is bounded as follows
\begin{equation}
 \frac{1}{2}\log{N}\leq S^{map}(\Phi_{self})\leq \log{N}.
\label{logi}
\end{equation}
\end{proposition}
The proof is given in Appendix~\ref{appentr}. The lower bound in (\ref{logi}) 
is not saturated as we show in Sec.~\ref{sectionSCN2}. 

Let us now estimate a classical channel capacity defined as a maximum rate in which classical information is transmitted through the channel. A formal definition (see for instance~\cite{Haiashi2006}) is analogous to Eq.~(\ref{channelcap1}), where now $d$ stands for the dimensionality of a vector space of bits strings transmitted through the channel with vanishing error. It has been shown~\cite{Holevo1998,Schumacher1997,Ogava1999} that the classical capacity $C_c$ of a quantum channel $\Phi$ can be expressed as
\begin{equation}
 C_c(\Phi)=\sup_{\{p_i,\rho_i\}_{i=1}^m} \chi\left(\{p_i,\Phi(\rho_i)\}_{i=1}^m\right),
 \label{cece}
\end{equation}
where $\{p_i\}_{i=1}^m$ is a probability density characterising a message of $m$ letters encoded in an alphabet of quantum states $\{\rho_i\}_{i=1}^m$. Here $\chi$ is the Holevo information defined by
\begin{equation}
 \chi(\{p_i,\rho_i\}_{i=1}^m)\equiv S\left(\sum_{i=1}^mp_i\rho_i\right)-\sum_{i=1}^mp_iS(\rho_i).
\end{equation}
A particular choice of the ensemble $\{p_i,\rho_i\}_{i=1}^m$ in $\chi(\{p_i,\Phi(\rho_i)\}_{i=1}^m)$ gives a lower bound on the classical capacity.  Let us consider $p_i=1/m$ for each $i$ and $\rho_i=|\phi_i\rangle$ such that $\langle\phi_j|\phi_i\rangle=\delta_{ij}$, where $\delta_{ij}$ is the Kronecker delta. Then one arrives at the following bound,
\begin{equation}
 C_c(\Phi)\geq S(\Phi(\rho_*))-\sum_{i=1}^mp_iS(\Phi(|\phi_i\rangle)).
 \label{ccbound}
\end{equation}
In Sec.~\ref{sectionSCN2} we show for single-qubit selfcomplementary channels that the term on the right hand side is usually strongly greater than zero. This implies that the classical capacity of a selfcomplementary channel is usually greater than zero.

\section{One-qubit family of selfcomplementary channels}\label{sectionSCN2}

In this section we find a parameterization of single-qubit selfcomplementary channels. Consider an arbitrary selfcomplementary channel $\Phi_{self} : {\cal{M}}_2 \to {\cal{M}}_2$. It can be defined by two Kraus operators
\begin{equation}
K_1=\begin{bmatrix} a_1&a_2\\a_3&a_4\end{bmatrix}, \quad \quad  K_2=\begin{bmatrix} b_1&b_2\\b_3&b_4\end{bmatrix}.
\end{equation}
In the set of selfcomplementary channels one can introduce a foliation of unitarily equivalent classes of channels, since a unitary transformation of an input state and a unitary transformation of output states of the channel and its complementary counterpart do not change the selfcomplementarity of the channel. In each equivalence class we have channels with Krauss operators related by $K_1'=WK_1V^{\dagger}$ and $K_2'=WK_2V^{\dagger}$, where $W$ and $V$ are arbitrary. Matrices $W$ and $V$ can be chosen in such a way that they transform the first Kraus operator to the diagonal form by the 
singular values decomposition, then
\begin{equation}
\label{diagprim}
K'_1=\begin{bmatrix} a&0\\0&b\end{bmatrix}, \quad \quad K'_2=\begin{bmatrix} c&d\\e&f\end{bmatrix},
\end{equation}
where $a$ and $b$ are non-negative numbers. Relation~(\ref{pantalon}) for selfcomplementary channels requires that the second row of the first Kraus operator has to be equal to the first row of the second one. This guarantees that the Kraus operators of the channel and its complementary counterpart are the same. Therefore the most general form of these operators for single-qubit selfcomplementary channels up to local unitary transformations takes the following form
\begin{equation}
K'_1=\begin{bmatrix} a&0\\0&b\end{bmatrix}, \quad \quad K'_2=\begin{bmatrix} 0&b\\\gamma&\delta\end{bmatrix}.
\end{equation}
Completeness relation, $\sum_{i=1}^{k} {K'}_{i}^{\dagger}{K'}_{i} = {\bf 1}$, imply 
additional constraints,
\begin{equation}
\begin{cases} \delta=0 ,\\ 2b^2=1,\\|\gamma|^2+a^2=1\end{cases}\qquad {\text or}\qquad \begin{cases} \gamma=0, \\ a^2=1,\\|\delta|^2+2b^2=1\end{cases},
\end{equation}
which allow us to reduce the number of parameters. These conditions imply that single-qubit selfcomplementary channels can be divided into two classes of maps,
 each of them characterized by two real parameters,
\begin{equation}
K'_1=\begin{bmatrix} {\sin \theta}&0\\0&{{\frac{1}{\sqrt{2}}}}\end{bmatrix},\qquad
K'_2=\begin{bmatrix} 0&{{\frac{1}{\sqrt{2}}}}\\ {\cos \theta e^{i\varphi}}&0\end{bmatrix} 
\label{OneQubitMap}
\end{equation}
or
\begin{equation}
K'_1=\begin{bmatrix} 1&0\\0&\frac{1}{\sqrt{2}}\sin\theta\end{bmatrix},\qquad
K'_2=\begin{bmatrix} 0&\frac{1}{\sqrt{2}}\sin\theta\\ 0&\cos{\theta}e^{i\varphi}\end{bmatrix}, 
\label{OneQubitMap1}
\end{equation}
where the free phases satisfy $\theta \in [0,\pi]$ and $\varphi\in [0,2\pi]$.
Furthermore, each selfcomplementary channel depends on two arbitrary unitary matrices
$V$ and $W$ used to bring $K_1$ to the diagonal form \ref{diagprim}.

Substituting $\theta=0$ and $\varphi=0$ in Eq.~(\ref{OneQubitMap1}), one gets the dephasing channel  that is characterized by maximum classical capacity but vanishing quantum capacity. Indeed, when the input state is either $|0\rangle$ or $|1\rangle$ the dephasing channel and its complementary act as a perfect classical copy machine. However, since all the coherences are destroyed, 
there are no coherences in any superposition of these quantum states.

Due to the Stinespring dilation theorem every quantum channel can be represented as a 
partial trace of an extended system subjected to a
global unitary transformation (\ref{EnvirnomentRepresentation}),
which couples the system with its environment.
 Let us now recall the unitary transformation corresponding to a selfcomplementary channel. 
A relation between a set of Kraus operators and the corresponding unitary transformation 
is shown in Appendix~\ref{trombalski}, see also~\cite{Zycz},
\begin{equation}
U_{ijk\nu}=\langle i | \otimes \langle j | U| k \rangle \otimes | \nu \rangle= {K'}_{ik}^j.
\label{unita}
\end{equation} 
Exact form of the global unitary operation $U$ for a single-qubit selfcomplementary channel represented by Eq.~(\ref{OneQubitMap}) reads
\begin{equation}
 U=\begin{bmatrix}\sin \theta&0&0&-\cos \theta e^{-i\varphi}\\0&1/\sqrt{2}&1/\sqrt{2}&0\\0&1/\sqrt{2}&-1/\sqrt{2}&0\\\cos \theta e^{i\varphi}&0&0&\sin \theta\end{bmatrix}.
\end{equation}

\begin{figure}[ht!]
  \begin{center}
    \scalebox{0.9}{\includegraphics[width=0.7\textwidth]{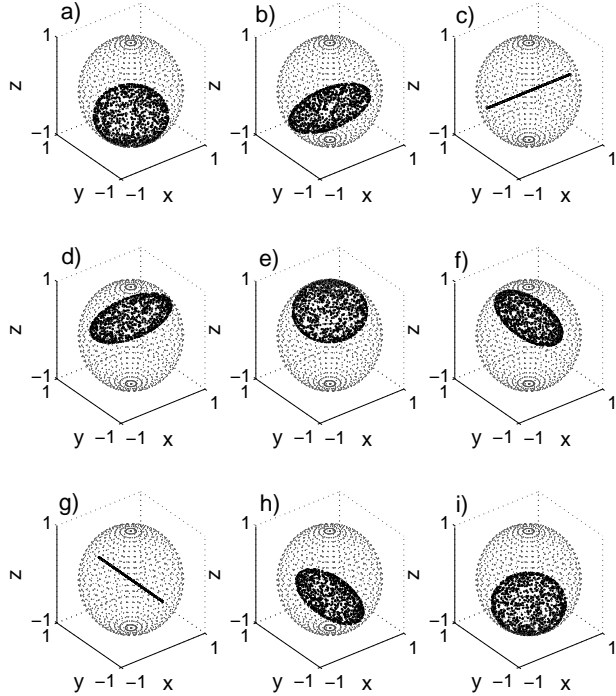}}
 \caption{Images of the Bloch sphere after the action of
 single-qubit selfcomplementary maps described in Eq.~(\ref{OneQubitMap}).
Figures $a)$ to $i)$ represent the channels for $\varphi=0$ and the main parameter reads $\theta=k \frac{\pi}{8}$ for $k=0,...,8$ respectively.}\label{SCstrucure}
  \end{center}
\end{figure}

Parameterization of selfcomplementary channels given in Eq.~(\ref{OneQubitMap}) or Eq.~(\ref{OneQubitMap1}) allows us to visualize the action of these channels on single-qubit pure states. Each density matrix representing a single-qubit state can be decomposed in the basis of the Pauli matrices 
and represented by a three-dimensional real vector of length $\leq 1$. 
These vectors form a ball called the Bloch ball, 
while vectors representing pure states of a single qubit form the Bloch sphere. 
An image of the Bloch sphere after an action of a  quantum channel 
 allows us to study decohering properties of the channel. Figure~\ref{SCstrucure} shows 
the action of different selfcomplementary channels given by~Eq.~(\ref{OneQubitMap}) 
for $\varphi=0$ and $\theta=k \frac{\pi}{8}$ for $k=0,...,8$. 
Among them, we can see the deformations of the Bloch sphere
 strong enough to make the quantum channel capacity equal to zero. 
Observe that the family of selfcomplementary maps does not include 
neither channels close to unitary nor maps close to the maximally depolarizing  channel.
Indeed, interaction with a two level environment cannot cause depolarizing of all states to a single  point inside the Bloch ball. 
Panels $c)$ and $g)$ show examples of decohering channels which cause 
projection of all the states into the line unitarily equivalent to the set of classical states.

Notice that one-qubit selfcomplementary channels are generically not bistochastic for $\theta=\{ 0, \pi \}$, which means that they do not preserve the maximally mixed state.
As it has been shown in Sec.~\ref{secentropy} the entropy of the image of the maximally mixed state gives us the entropy of the selfcomplementary channel, 
$S^{map}(\Phi^{self})=S\left(\Phi(\rho_*)\right)$. 
Fig.~\ref{SCstrucure} shows that the entropy of the image of maximally mixed state cannot be arbitrary small. Its minimum reads 
$S\left(\left[\frac{1}{4},\frac{3}{4}\right]\right) \simeq 0.56233$, see panels $a)$, $e)$ and $i)$, which does not saturate the lower limit provided by Proposition~\ref{propent}. 

\begin{figure}[ht!]
  \begin{center}
    \scalebox{0.5}{\includegraphics[width=1\textwidth]{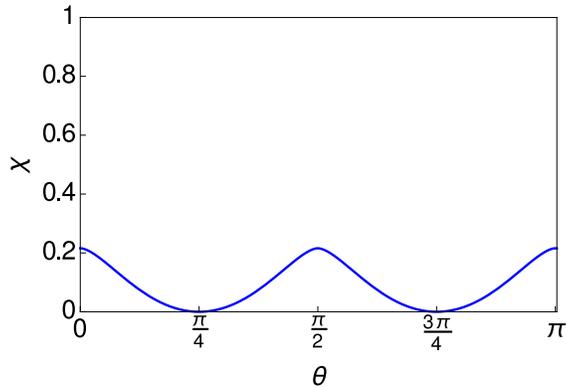} }
 \caption{Lower bound of the classical channel capacity for one qubit selfcomplementary channels given in Eq.~(\ref{OneQubitMap}) with $\varphi=0$ as a function of  the phase $\theta$. 
The lower bound is given by the Holevo information $\chi$ defined in the r.h.s. of Ineq.~(\ref{ccbound})  for states $|0\rangle$ and $|1\rangle$  occurring with equal probabilities.}
\label{chi}
  \end{center}
\end{figure}

The exact parameterization of one-qubit selfcomplementary channels allows us to find exact entropies for the output states. This, in turn, allows us to estimate
the lower bound on the classical capacity given in Ineq.~(\ref{ccbound}). For selfcomplementary channels characterized by Eq.~(\ref{OneQubitMap}) with $\varphi=0$ the lower bound on this capacity is plotted in Fig.~\ref{chi}.

This figure shows that the classical capacity for selfcomplementary channels is significantly greater than zero. Therefore, these channels although noisy enough to have quantum channel capacity equal to zero, are not completely closed for transmitting of the classical information. 
The coherences, although strongly weakened, are also not entirely destroyed. 
This suggests that also some residual entanglement can be preserved by selfcomplementary channels. 
This problem is analyzed in the following section. 

\section{Residual entanglement preserved by selfcomplementary channels}\label{resent}

In this section, we analyze two measures of entanglement of the Choi-Jamio{\l}kowski states corresponding to selfcomplementary channels in order to show that these channels are not generically entanglement breaking. A quantum channel is entanglement breaking if acting locally on a part of an entangled state produces an output which is not entangled with the remaining part independently of the initial state.  It is known \cite{Horodecki2003} that a channel is entanglement breaking if and only if the corresponding Choi-Jamio{\l}kowski state defined in Eq.~(\ref{choijam}) is separable. 
\vspace{+1.0cm}
\begin{figure}[ht!]
  \begin{center}
    \scalebox{0.7}{\includegraphics[width=1.2\textwidth]{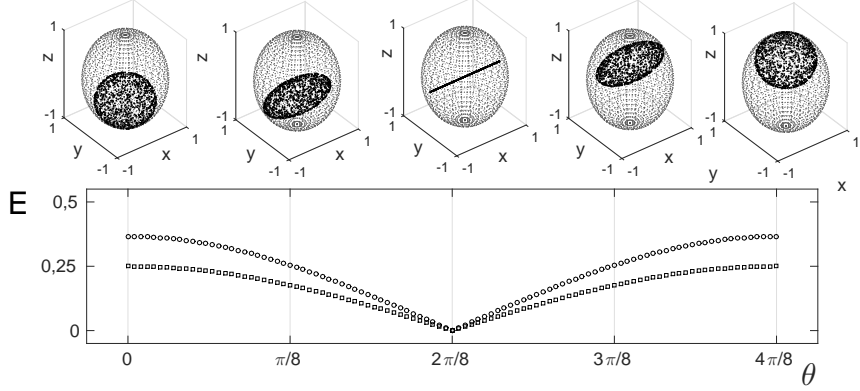}}
 \caption{One-qubit selfcomplementary maps (\ref{OneQubitMap}) and entanglement $E$ of the 
  corresponding Choi-Jamio{\l}kowski states.
  Figures $a)$ to $e)$ represent the images of the Bloch sphere induced by the consecutive channels
   obtained for $\varphi=0$ and the phase $\theta=k \frac{\pi}{4}$, with $k=0,...,4$.
   Negativity (squares) and concurrence (circles) 
   of the corresponding states $D_{\Phi}/N$ is  shown in the lower panel
   as functions  (\ref{NegFunction}) and (\ref{ConcurrenceSingleQubit})
    of the phase $\theta$, respectively.
}
\label{NegFig}
  \end{center}
\end{figure}

\begin{figure}[ht!]
  \begin{center}
    \scalebox{0.9}{\includegraphics[width=0.8\textwidth]{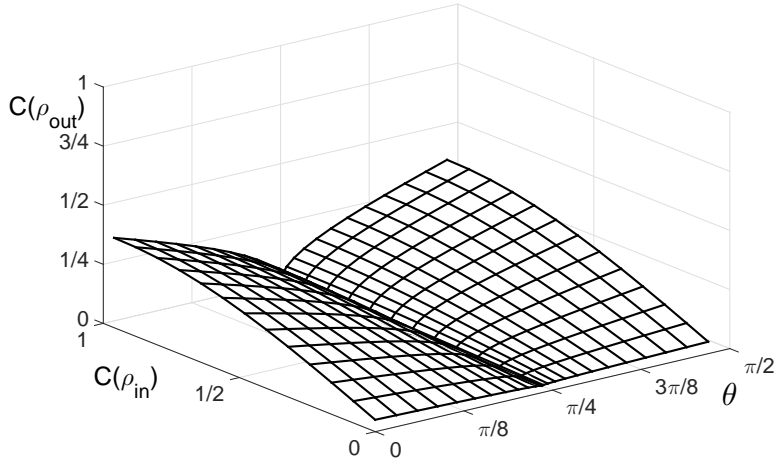} }
 \caption{Entanglement evolution measured by concurrence $C$ of the family of single-qubit selfcomplementary maps defined in Eq.~(\ref{OneQubitMap}), where $\varphi=0$ 
and the phase $\theta \in [0,\pi/2]$. 
Here $C(\rho_{in})$ and $C(\rho_{out})$ denote 
concurrence of the input state and the output state respectively.}
\label{EntanEvol}
  \end{center}
\end{figure}

Let us analyze entanglement of a Choi-Jamio{\l}kowski state  $\omega_\Phi$ of a single-qubit selfcomplementary channel $\Phi$. For the channel given in Eq.~(\ref{OneQubitMap}) with $\varphi=0$ we have
\begin{equation*}
 \omega_\Phi=\frac{D_{\Phi}}{2}=\frac{1}{2}\begin{bmatrix} \sin^2\theta&0&0&\frac{1}{\sqrt{2}}\sin\theta\\0&\frac{1}{2}&\frac{1}{\sqrt{2}}\cos\theta&0\\0&\frac{1}{\sqrt{2}}\cos\theta&\cos^2\theta&0\\\frac{1}{\sqrt{2}}\sin\theta&0&0&\frac{1}{2}\end{bmatrix},
\end{equation*}
where we have applied a relation between the Kraus representation and the Choi-Jamio{\l}kowski state discussed in Appendix~\ref{trombalski}.
As a measure of entanglement we take an entanglement monotone called the {\sl concurrence}~\cite{Hill1997}. For a two-qubit mixed state $\omega_\Phi$ it is defined as 
\begin{equation}
C(\omega_\Phi)=max\{0, \sqrt{\gamma_1}-\sqrt{\gamma_2}-\sqrt{\gamma_3}-\sqrt{\gamma_4} \},
\label{Concurrence}
\end{equation}
where the $\gamma_1 \le \gamma_2 \le \gamma_3 \le \gamma_4$ are the eigenvalues of $R=\omega_\Phi \tilde{\omega}_\Phi$. Here $\tilde{\omega}_\Phi$ is the result of a spin-flip operation applied to $\omega_\Phi$:                  
\begin{equation}
\tilde{\omega}_\Phi=(\sigma_y \otimes \sigma_y)\omega_\Phi^{*}(\sigma_y \otimes \sigma_y)
\end{equation}              
and the complex conjugation is taken in the computational basis. Explicit formula for the concurrence for arbitrary two-qubit states has been found by Wootters~\cite{Wootters}. The concurrence of the Choi-Jamio{\l}kowski state of a single-qubit channel plays the role of the proportionality factor in a relation between entanglement of an input and an output state~\cite{Buchleitner},
\begin{equation}
C(\rho_{out})=C(\rho_{in})C(\omega_\Phi(\theta)).
\label{EntanglementEvolution}
\end{equation}
This allows us to characterize residual entanglement remaining after transformation driven by selfcomplementary channels. For these channels parametrized  as in Eq.~(\ref{OneQubitMap}) with $\varphi=0$ the matrix $R$ reads
\begin{equation}
  R=\omega_\Phi \tilde{\omega}_\Phi=\frac{1}{2}\begin{bmatrix} \frac{1}{2}&0&0&\frac{1}{\sqrt{2}}\cos\theta\\0&\frac{1}{2}&\frac{1}{\sqrt{2}}\sin\theta&0\\0&\frac{1}{\sqrt{2}}\sin\theta&\cos^2\theta&0\\\frac{1}{\sqrt{2}}\cos\theta&0&0&\sin^2\theta\end{bmatrix}.
\end{equation}
Its eigenvalues are given by
\begin{equation}
\gamma_1=-\frac{1}{4} (4\cos2\theta-1),\qquad \gamma_2=-\frac{1}{4} (2-\cos2\theta),\qquad \gamma_3=0,\qquad \gamma_4=0.
\label{NegEigenValues}
\end{equation}
so the concurrence reads,
\begin{equation}
C(\omega) =\begin{cases}
   \frac{1}{2} ( \sqrt{4\cos2\theta-1}-\sqrt{2-\cos2\theta} ), \qquad \theta \in [0;\frac{\pi}{4})\\ 
    0, \qquad \qquad \qquad \qquad \qquad \quad\qquad \qquad \theta=\frac{\pi}{4}\\
    \frac{1}{2} ( \sqrt{2-\cos2\theta}-\sqrt{4\cos2\theta-1} ), \qquad  \theta \in (\frac{\pi}{4};\frac{\pi}{2}]
   \end{cases}.
\label{ConcurrenceSingleQubit}
 \end{equation} 
Figure~\ref{NegFig} shows that selfcomplementary channels preserve residual entanglement of the initial maximally entangled state, if only a single part of this state is transformed by one of these channels. Only the linear channel related to $\theta=\pi/4$, is entanglement breaking, as 
the corresponding state is separable -- see Fig.~\ref{NegFig}. 
The maximum concurrence is achieved  for the amplitude damping channel, $\theta=\pi/2$ defined by the following Kraus operators $K_{AD}$ 
\begin{equation}
K_1^{(AD)}=\begin{bmatrix} 1&0\\0&\sqrt{1-p}\end{bmatrix},\qquad
K_2^{(AD)}=\begin{bmatrix} 0&\sqrt{p}\\0&0\end{bmatrix}
\label{N2amplitudedamping}
\end{equation}          
where $p$ indicates a probability of decaying to the ground state. 
Dependence of the entanglement of the output state on the entanglement of an input state for selfcomplementary channels is shown in Fig.~\ref{EntanEvol}. It is evident that almost all single-qubit selfcomplementary channels preserve some residual entanglement.

The concurrence is a measure of entanglement characterizing two-qubit states. As a measure that can be applied also for larger quantum systems we take an entanglement monotone called {\sl negativity}~\cite{Zyczkowski1998, Eisert2001, Vidal2002} defined as follows
\begin{equation}
 Neg(\omega_\Phi)=\frac{{||\omega_\Phi^{T_A}||}_1-1}{2},
\label{NegativitySys}
\end{equation}
where the partial transpose ${T_A}$ with respect to subsystem $A$ is defined as
\begin{equation}
  \omega_\Phi^{T_A}=\sum_{ijkl} p_{ijkl} (|i\rangle \langle j|)^T \otimes |k\rangle \langle l|=\sum_{ijkl} p_{ijkl} |j\rangle \langle i| \otimes |k\rangle \langle l|.
\end{equation} 
 Straightforward calculations lead us to the following formula for negativity of the Choi-Jamio{\l}kowski state of the single-qubit selfcomplementary channels 
\begin{equation}
N_{\omega_\Phi}(\theta)=\frac{1}{4} |\cos2\theta|.
\label{NegFunction}
\end{equation}
As shown in Fig. \ref{NegFig}, 
both measures of entanglement satisfy inequality $C \ge N$,
originally observed in \cite{Zy99}.

Let us also notice that for single-qubit selfcomplementary channels the concurrence is a monotonic function of the negativity 
\begin{equation}
C({\omega_\Phi}) =\begin{cases} 
   \frac{1}{2} ( \sqrt{16N_{\omega_\Phi}(\theta)-1}-\sqrt{2-4N_{\omega_\Phi}(\theta)} ), \qquad \theta \in [0;\frac{\pi}{4})\\ 
    0, \qquad \qquad \qquad \qquad\quad \ \qquad \qquad \qquad \qquad \theta=\frac{\pi}{4}\\
    \frac{1}{2} ( \sqrt{2-4N_{\omega_\Phi}(\theta)}-\sqrt{16N_{\omega_\Phi}(\theta)-1} ), \ \ \ \quad \theta \in (\frac{\pi}{4};\frac{\pi}{2}]
   \end{cases}.
\label{ConcurrenceSingleQubit2}
 \end{equation}

\subsection{Selfcomplementary dynamics, characterization of non-Markovianity}

A global unitary transformation considered in Eq.~(\ref{unita}) that provides a coupling of a qubit system with a qubit environment can represent a selfcomplementary dynamics,
if we assume that the phase changes linearly with time, $\theta=\omega t$.
In such a dynamics, information oscillates between the system and the environment and the evolution depends on the history. Fig.~\ref{SCstrucure} provides an illustration of this process. The successive images of the Bloch spheres represent now the successive moments of time. 
In panel $c)$, the Bloch ball is contracted to a line segment. 
Then the points diverge to form a three-dimensional set again. 
This evolution clearly depends on both the present state and the previous history. This type of memory-based processes is called non-Markovian. In contrast, the 
so--called Markovian dynamics depends only on the present state of the quantum system.

By Stinespring dilation theorem represented by Eq.~(\ref{EnvirnomentRepresentation}), every completely positive and trace-preserving (CPTP) map - a quantum channel - can be described by an interaction with an environment in a pure state. Therefore, the dynamics induced by such a channel does not depend on the previous evolution of an input state, but only on the actual state of this system. In consequence, if a Markovian evolution is described by a CPTP quantum channel
then it can be decomposed into a concatenation of infinitely many CPTP maps. 
Each of them can be represented by an interaction with an independent environment according to Eq.~(\ref{EnvirnomentRepresentation}) and
 each part of the evolution removes the information about the previous evolution of the input state. 
On the other hand, if a process cannot be decomposed into infinitesimal 
CPTP maps then it is non-Markovian.

Recently, many efforts have been made to recognize and characterize non-Markovianity of a
quantum evolution. One of witnesses of the non-Markovianity is 
based on the observation that certain quantities, as  quantum channel capacity~\cite{Bylicka2014}, decrease monotonically for concatenation of CPTP maps. 
Therefore, if during an evolution one observes an increase of the channel capacity,
such a process is non-Markovian, 
 and the evolution cannot be described as a concatenation of infinitesimal CPTP maps. 
 However, such the non-monotonic behavior of the channel capacity provides 
 a sufficient but not necessary condition for the non-Markovianity.  
 Indeed, the selfcomplementary evolution provides an example of highly non-Markovian 
 dynamics, for which the quantum channel capacity is always zero.

In this case, a better characteristic of the non-Markovianity is given by the changes of the
entanglement of the Choi-Jamio{\l}kowski state shown in Fig.~\ref{NegFig}. The entanglement cannot increase under concatenation of CPTP maps. Since we observe that entanglement increases 
during this process, this evolution is non-Markovian. 
A degree of non-Markovianity can be characterized by the sum of all time intervals over which the entanglement increases. During the selfcomplementary evolution this measure is infinite, since information oscillates between system and environment without being damped. 

\section{Concluding remarks}\label{sec:remarks}

In the present paper we investigated a class of selfcomplementary 
quantum channels which send the state of the principal system and the state of the
environment into the same output state.
In the simplest case we characterize the selfcomplementary maps of a single qubit, which
up to local unitary operations are parameterized  by two real phases 
- see Eq.~(\ref{OneQubitMap}) and Eq.~(\ref{OneQubitMap1}). 
A  generalization of this parameterization for higher dimensions is provided in Appendices~\ref{gutritgen} and~\ref{sectionSCNall}.

Furthermore, we analyzed classical and quantum capacities of single-qubit selfcomplementary channels. Moreover, we studies decoherence and changes of entanglement, they induce. 
Two possible ways to interpret the concurrence of the Choi-Jamio{\l}kowski state related to a single-qubit selfcomplementary channel are proposed.
On one hand, the concurrence is  a proportionality factor in a relation between the 
concurrences of an input and an output state, where only one part of the system
is transmitted through the channel. On the other hand, the changes of the concurrence characterize non-Markovian character of an evolution given by a family of selfcomplementary channels.

Since a selfcomplementary channel transforms a quantum state into two identical states of the system and the environment, such a map describes an approximate quantum copying machine.
The machine is not perfect due to the no-cloning theorem, which 
implies that the multiplied states are generically  different from the initial state. 
This theorem additionally implies zero capacity of the selfcomplementary channels. 
Indeed, if there had been a Hilbert subspace from which all the states are transmitted with arbitrary high fidelity, then multiplication induced by the selfcomplementary channels would have caused a violation of the no-cloning theorem for the entire subspace.

Selfcomplementary channels appear in wider context as particular examples of the so called {\sl degradable } and {\sl anti-degradable} channels~\cite{Wilde,Lloyd,Shor}. A channel $\Phi$  is called degradable if there exists another completely positive trace preserving map $\Psi$ such that 
\begin{equation}
\Psi \circ \Phi=\widetilde{\Phi},
\label{DegradableCond}
\end{equation}
where $\widetilde{\Phi}$ is the complementary channel. A channel $\Phi$ is called {\sl anti-degradable} if its complementary channel $\widetilde{\Phi}$ is degradable with respect to original quantum system and satisfies the relation
\begin{equation}
  \Psi \circ \widetilde{\Phi}=\Phi.
\label{AntiDegradableCond}
\end{equation}
The argument derived from the no-cloning theorem implies that all anti-degradable channels have zero quantum channel capacity~\cite{Lloyd,Ruskai}.
In~\cite{Ruskai} single-qubit degradable channels are completely characterized and it is shown that single-qubit channels with two Kraus operators are either degradable or anti degradable, see also~\cite{Wolf2007}.   

The classical capacity of selfcomplementary channels is not zero. Our results show also that these channels do not destroy completely neither the coherences nor the entanglement. These features allow us to pose a question, whether 
the quantum capacity of the selfcomplementary maps could be activated by other zero capacity channel
 if the two channels act in  parallel. 
This kind of superactivation of two zero capacity quantum channels has been observed previously, see for instance~\cite{Smith}. Our analysis implies that in order to superactivate a selfcomplementary channel the second channel cannot be selfcomplementary. 

Notice that the activation of a selfcomplementary channel by another channel does not violate the no-cloning theorem. Indeed, the second channel does not copy the corresponding system to its environment. The joint state of the environments is no longer a copy of the joint output state. Therefore, without violating the no cloning theorem, the output state could in principle be
similar to the input state as far as the joint state of the environments is different from them. 
This may hold despite the similarity of partial states of output 
from the selfcomplementary channel and the corresponding partial state of the environment.  
However, further investigation on possible activation of selfcomplementary maps
is still required.



\section*{Acknowledgments.}
It is a pleasure to thank Pawe{\l} Horodecki for numerous 
discussions and valuable remarks.
Financial support by the project \#56033 financed by the Templeton Foundation 
and by the grants financed by the Polish National Science Center under the contracts number
2011/03/N/ST2/01968 (MS) and DEC-2011/02/A/ST1/00119 (K{\.Z})  is gratefully acknowledged.

\appendix

\section{Quantum channels and their representations}\label{trombalski}

In this Appendix, we review the formalism of quantum channels 
 used in the main body of the paper and in the proofs of the propositions provided in other Appendices.
  A quantum map $\Phi:\rho\rightarrow \rho'$ that describes an interaction of a 
  quantum system $\rho$ with an environment can be represented as completely positive, 
 and trace-preserving (CPTP) transformation~\cite{Jam,Choi,Kraus,Pillis}.
{\sl Complete positivity}  means that an extended map $\Phi\otimes\1_M$, where $\1_M$ denotes an identity operator acting on $M$ dimensional space of density matrices, preserves positivity of the matrices for any $M$. Completely positive and trace preserving quantum maps are called 
 {\sl quantum operations}, {\sl stochastic maps} or {\sl quantum channels}. 


Due to the theorems of Jamio{\l}kowski \cite{Jam} and Choi \cite{Choi} 
  complete positivity of a map is equivalent to positivity of a state 
corresponding to the map by the {\sl Jamio{\l}kowski isomorphism}. This isomorphism determines the correspondence between a quantum operation $\Phi$ acting on $N$ dimensional matrices and density matrix $D_{\Phi}/N$ of dimension $N^2$ which is called the Choi-Jamio{\l}kowski state and is defined as follows
\begin{equation}
\frac{1}{N}D_{\Phi}=[\1_N\otimes\Phi]\big(|\phi^+\left.\right\rangle\left\langle\right. \phi^+|\big),
\label{miopio}
\end{equation}
where $|\phi^+\left.\right\rangle=\frac{1}{\sqrt{N}}\sum_{i=1}^N |i\left.\right\rangle\otimes|i\left.\right\rangle$ is the maximally entangled state. The 
Choi matrix $D_{\Phi}$ corresponding to a trace preserving operation satisfies the following condition 
\begin{equation}
{\rm Tr}_{2} D_\Phi=\1,
\label{partialtrace}
\end{equation}
where ${\rm Tr}_{2}$ is a partial trace over the second subsystem of the state in Eq.~(\ref{miopio}).

A quantum operation $\Phi$ can also be represented by a {\sl superoperator matrix}. 
It is a matrix that acts on a vector of length $N^2$ containing all the entries
of the density matrix $\rho_{ij}$ of an input state ordered lexicographically. 
Thus, the superoperator of $\Phi$ is represented by a square matrix of size $N^2$.
The superoperator in some orthogonal product basis $\{|i \rangle \otimes |j \rangle \}$ is represented by a matrix indexed by  four indexes, 
\begin{equation}
\Phi_{ij,kl}=\langle i | \otimes \langle j | \Phi | k \rangle \otimes | l \rangle.
\label{zy1}
\end{equation} 
The matrix from Eq.~(\ref{miopio}) represented in the same basis is related to the superoperator matrix by a reshuffling formula \cite{Zycz} as follows
\begin{equation}
\langle i | \otimes \langle j | D_{\Phi} | k \rangle \otimes |l \rangle = \langle i | \otimes \langle k | \Phi | j \rangle \otimes | l \rangle.
  \label{zyrafa}
\end{equation}
The entropy of (\ref{miopio}) is called the map entropy and denoted as $S^{map}(\Phi)$,
\begin{equation} \label{SCentropy}
S^{map}(\Phi)\equiv S\left(\frac{1}{N}D_{\Phi}\right)=S\left([1_N\otimes\Phi]\big(|\phi^+\left.\right\rangle\left\langle\right. \phi^+|\big)\right). 
\end{equation}

To describe a quantum channel, one may use the Stinespring's dilation theorem \cite{Stinespring1955} concerning an initial state $\rho$ on ${\mathcal H}_N$, interacting with its environment characterized by a state on ${\mathcal{H}}_M$. The joint evolution of the two states is described by a unitary operation $U$. The joint state of the system and the environment is initially not entangled. Moreover, the initial state of the environment can be given by a pure one without lost of generality. The evolving joint state is given by
\begin{equation}
\omega=U\Big(\left|1\right\rangle\left\langle 1\right|\otimes \rho \Big)U^{\dagger},
\label{omom}
\end{equation}
where $|1\rangle \in {\cal{H}}_M$ and $U$ is a unitary matrix of size $NM$. The state of the system  after the operation is obtained by tracing out the environment, 
\begin{equation}
\rho'=\Phi(\rho)={\rm Tr}_M \Big[U\big(\left|1\right\rangle\left\langle 1\right|\otimes \rho\big) U^{\dagger}\Big]=\sum_{i=1}^M K^i\rho K^{i\dagger},
\label{quantop}
\end{equation}
where the so called Kraus operators $K^i$ read $K^i=\left\langle i\right| U\left|1\right\rangle$. The Kraus operators $\{K^i\}$ satisfies completeness relation
 $\sum_{i=1}^{k} K_{i}^{\dagger}K_{i} = \1$ that implies preservation of positivity. In the matrix representation the Kraus operators are formed by successive
blocks of the first block--column of the unitary evolution matrix $U$. 
Due to the Kraus theorem~\cite{Kraus} a map $\Phi$ is completely positive if and only if there exists a Kraus representation
\begin{equation}
\rho'=\Phi(\rho)=\sum_{i=1}^M K^i\rho K^{i\dagger}.
\label{mrowkojad}
\end{equation}
A superoperator matrix is related to the Kraus operators by the following formula
\begin{equation}
{\Phi=\sum_{i=1}^{k}} K_{i} \otimes \overline{K_{i}}. 
\label{KrausTensor}
\end{equation}
This relation together with Eq.~(\ref{zyrafa}) allow us to express the Choi-Jamio{\l}kowski state by the Kraus operators.



\section{Proof of Proposition~\ref{propkraus}}\label{appkraus}
This proposition concerns a relation between Kraus operators of a quantum channel and its complementary counterpart. The Stinespring's dilation theorem allows us to express a channel in the following form $\Phi(\rho_s)={\rm Tr}_s[U(|1_e\rangle\langle 1_e|\otimes \rho_s )U^{\dagger}]$, where $s$ denotes the system and $e$ the environment. This formula can be written by using the swap operator $O_{SWAP}$ which exchanges the system and the environment as follows ${\rm Tr}_e[O_{SWAP} U(|1_e\rangle\langle 1_e|\otimes \rho_s )U^{\dagger}O_{SWAP}^{\dagger}]$. Since the Kraus operators of the channel read $K^i_{jk}=\left\langle ij\right| U\left|1k\right\rangle$, we have the Kraus operators of the complementary counterpart given by
\begin{equation}
\widetilde{K}^i_{jk}=\left\langle ij\right|O_{SWAP} U\left|1k\right\rangle=\left\langle ji\right|U\left|1k\right\rangle=K^j_{ik}.
\end{equation}
This justifies Proposition~\ref{propkraus}.


\section{Proof of Proposition~\ref{PropentTensor}}\label{appproduct}
This proposition concerns the fact that the tensor product of selfcomplementary channels is also selfcomplementary. Let us consider two selfcomplementary channels $\Psi_Z$ and $\Psi_R$, which are characterized according to (\ref{mrowkojad}) by sets of the Kraus operators $\{Z^i\}$ and $\{R^j\}$, respectively. Both the sets of satisfy Eq.~(\ref{pantalon}). One can demonstrate that the Kraus operators of $\Psi_Z \otimes \Psi_R$, which are $K^{ij}=Z^i \otimes R^j$ satisfy Eq.~(\ref{pantalon}) as well. In what follows, the lower indexes represent the matrix elements according to the same convention as in Eq.~(\ref{zy1}),
\begin{align}
 K_{psrt}^{ij}
  &={[Z^i \otimes R^j]}_{psrt}=Z_{pr}^i R_{st}^j=Z_{ir}^p R_{jt}^s \\
  &={[Z^p \otimes R^s]}_{ijrt}=K_{ijrt}^{ps}.
\end{align}
This proves that the tensor products preserves relation~(\ref{pantalon}) and justifies Proposition~\ref{PropentTensor}.

\section{Proof of Proposition~\ref{propent}}\label{appentr}

Proof of part $a)$ concerning equivalence between the map entropy and the output entropy for selfcomplementary channels if the input state if maximally mixed. Let us construct a three-tripartite pure state $\rho_{ABC}$ on a Hilbert space ${\cal{H}}_{ABC}$, such that
\begin{equation}
 \rho_{ABC}={|\psi_+\rangle}{\langle \psi_+|}_{AB} \otimes {|1\rangle \langle 1|}_C,
\end{equation}
where ${|\psi_+\rangle}_{AB}$ is a maximally entangled states on $ {\cal{H}}_{AB}$ and ${| 1 \rangle}_C$ is a pure state of an environment $C$. Consider an action of a unitary transformation on subsystems $BC$. After this operation the state $\rho_{ABC}$ is transformed into a state $\rho_{ABC}'$ which is also pure. The partial traces of $\rho_{ABC}'$ are also marked by sign prim $'$. The partial trace over $AC$ describes a quantum channel, while the partial trace over $AB$ describes its complementary. For a selfcomplementary channel $\Phi_{self}$ we have the following equality 
between partial traces of $\rho_{ABC}'$,
\begin{equation}
 \rho_{B}'=\rho_{C}'.
\label{parteq}
\end{equation}
Since the state $\rho_{ABC}'$ is pure its complementary partial traces have the same entropies, 
\begin{equation}
S(\rho_{C}')=S(\rho_{AB}').
\label{compeq}
\end{equation}
The entropy $S(\rho_{AB}')$ is equal to the map entropy $S^{map}(\Phi_{self})$ by construction. Notice that $S(\rho_{B}')=S\left(\Phi_{self}(\rho_*)\right)$.
Due to Eqs.~(\ref{parteq}) and~(\ref{compeq}) the proof of part $a)$ is completed. 

Proof of part $b)$ concerning bounds on the map entropy for selfcomplementary channels. The right inequality of~(\ref{logi}) is implied by the fact that the dimensionality of an environment involved in a selfcomplementary transformation is equal to the dimensionality of the output state. The left inequality of~(\ref{logi}) is proved by using the triangle inequality (Araki--Lieb inequality), that states that for any bi--bipartite state $\rho_{XY}$ the following entropic inequality holds
\begin{equation}
|S(\rho_{X})-S(\rho_{Y})|\leq S(\rho_{XY}).
\end{equation}
We can apply the above inequality to the state $\rho_{AB}'$ constructed as in the proof of part $a)$. Notice that for selfcomplementary channels $S(\rho_{AB}')=S^{map}(\Phi_{self})=S(\rho_{C}')=S(\rho_{B}')$ and $S(\rho_{A}')=S(\rho_{*})=\log{N}$. Using Araki--Lieb inequality we obtain
\begin{equation}
S(\rho_A')\leq S(\rho_{AB}')+S(\rho_B') 
\end{equation}
which implies that 
\begin{equation}
\log{N}\leq 2S^{map}(\Phi_{self}).
\end{equation}
This completes the proof of Proposition~\ref{propent}.

\section{A family of single-qutrit selfcomplementary channels }\label{gutritgen}

Let us discuss the parameterization of single-qutrit selfcomplementary channels. A selfcomplementary channel $\Phi=\widetilde{\Phi} : {\cal{M}}_3 \to {\cal{M}}_3 $ is described by three Kraus operators
\begin{equation}
  K_1=\begin{bmatrix} a_{11}&a_{12}&a_{13}\\a_{21}&a_{22}&a_{23}\\a_{31}&a_{32}&a_{33}\end{bmatrix},
\quad 
  K_2=\begin{bmatrix} b_{11}&b_{12}&b_{13}\\b_{21}&b_{22}&b_{23}\\b_{31}&b_{32}&b_{33}\end{bmatrix},
\quad 
  K_3=\begin{bmatrix} c_{11}&c_{12}&c_{13}\\c_{21}&c_{22}&c_{23}\\c_{31}&c_{32}&c_{33}\end{bmatrix}.
\end{equation}
In the set of all such channels one can introduce the foliation of unitary equivalent classes of maps. In one such class $ K_i'=UK_iV^{\dagger}$, where $i=1,2,3$, 
while $U$ and $V$ are arbitrary $3 \times 3$ unitary matrices
determined by the singular value decomposition of  $K'_1=UK_1V$. 
This transformation brings the first Kraus operator to the diagonal form with non-negative entries,
so that
\begin{equation}
  K_1'=\begin{bmatrix} \alpha_{1}&0&0\\0&\alpha_{2}&0\\0&0&\alpha_{3}\end{bmatrix},
\quad
  K_2'=\begin{bmatrix} \beta_{11}&\beta_{12}&\beta_{13}
\\ \beta_{21}&\beta_{22}&\beta_{23}\\\beta_{31}&\beta_{32}&\beta_{33}\end{bmatrix},
\quad
  K_3'=\begin{bmatrix} \gamma_{11}&\gamma_{12}&\gamma_{13}\\
  \gamma_{21}&\gamma_{22}&\gamma_{23}\\\gamma_{31}&\gamma_{32}&\gamma_{33}\end{bmatrix}.
\end{equation}
The relation $\widetilde{K^i}_{\alpha j} = {K^{\alpha}}_{ij}$ implies that the Kraus operators take the form 
\begin{equation}
  K_1'=\begin{bmatrix} \alpha_{1}&0&0\\0&\alpha_{2}&0\\0&0&\alpha_{3}\end{bmatrix},
\quad
  K_2'=\begin{bmatrix}
 0&\alpha_{2}&0\\\beta_{21}&\beta_{22}&\beta_{23}\\\beta_{31}&\beta_{32}&\beta_{33}\end{bmatrix},
\quad
  K_3'=\begin{bmatrix}
 0&0&\alpha_{3}\\\gamma_{21}&\gamma_{22}&\gamma_{23}\\\gamma_{31}&\gamma_{32}&\gamma_{33}\end{bmatrix}.
\end{equation}
One can parameterize the Kraus operators by introducing a parameter $\theta$ and set of parameters given by an auxiliary unitary $3\times 3$ matrix 
$$ W=\begin{bmatrix} W_{11}&W_{12}&W_{13}\\W_{21}&W_{22}&W_{23}\\W_{31}&W_{32}&W_{33}\end{bmatrix}. $$ Let us introduce a rescaled unitary matrix $X=sW$ with $s\leq 1$, such that $XX^{\dagger}=s^2 \1_3$. The relation $ \sum_{i=1}^{k} {K'}_{i}^{\dagger}{K'}_{i} = {\bf 1} $ allows us to reduce the number of parameters. The structure of the Kraus operators is the following 
\begin{equation}
\begin{cases}
  K_1'=\begin{bmatrix} \cos\theta&0&0\\0&\frac{1}{\sqrt{2}}\cos\theta&0\\0&0&\frac{1}{\sqrt{2}}\cos\theta\end{bmatrix},\\
  K_2'=\begin{bmatrix} 0&\frac{1}{\sqrt{2}}\cos\theta&0\\W_{11}\sin\theta&W_{21}\sin\theta&W_{31}\sin\theta\\\frac{1}{\sqrt{2}}W_{22}\sin\theta&\frac{1}{\sqrt{2}}W_{12}\sin\theta&\frac{1}{\sqrt{2}}W_{23}\sin\theta\end{bmatrix},\\
  K_3'=\begin{bmatrix} 0&0&\frac{1}{\sqrt{2}}\cos\theta\\\frac{1}{\sqrt{2}}W_{22}\sin\theta&\frac{1}{\sqrt{2}}W_{12}\sin\theta&\frac{1}{\sqrt{2}}W_{23}\sin\theta\\W_{13}\sin\theta&W_{23}\sin\theta&W_{33}\sin\theta\end{bmatrix}.
\end{cases}
\end{equation}

\section{Selfcomplementary channels on arbitrary quantum systems}\label{sectionSCNall}

Consider now an $N$-dimensional selfcomplementary channel, 
 $\Phi=\widetilde{\Phi} : {\cal{M}}_N \to {\cal{M}}_N $. 
It can be specified by $N$ Kraus operators
\begin{equation}
\overbrace{
\begin{bmatrix}a_{11}&a_{12}&\ldots&a_{1N}\\a_{21}&a_{22}&\ldots&a_{2N}\\\vdots&\vdots&\ddots&\vdots\\a_{N1}&a_{N2}&\ldots&a_{NN}\end{bmatrix},
...,
\begin{bmatrix}z_{11}&z_{12}&\ldots&z_{1N}\\z_{21}&z_{22}&\ldots&z_{2N}\\\vdots&\vdots&\ddots&\vdots\\z_{N1}&z_{N2}&\ldots&z_{NN}\end{bmatrix}}^N .
\end{equation}
In the set of these channels one can introduce foliation of unitary equivalent classes of channels. In one such class $K_i'=UK_iV^{\dagger}$ where $U$ and $V$ can be arbitrary $N \times N$ unitary matrices. Assume that $U$ and $V$ transform the first Kraus operator into diagonal matrix 
by the singular value decomposition, so that
\begin{equation}
\overbrace{
\begin{bmatrix}\alpha_{1}&0&\ldots&0\\0&\alpha_{2}&\ldots&0\\\vdots&\vdots&\ddots&\vdots\\0&0&\ldots&\alpha_{N}\end{bmatrix},
...,
\begin{bmatrix}\omega_{11}&\omega_{12}&\ldots&\omega_{1N}\\\omega_{21}&\omega_{22}&\ldots&\omega_{2N}\\\vdots&\vdots&\ddots&\vdots\\\omega_{N1}&\omega_{N2}&\ldots&\omega_{NN}\end{bmatrix}}^N.
\end{equation}
The relation $\widetilde{K^i}_{\alpha j} = {K^{\alpha}}_{ij}$ 
implies further constraints on the Kraus operators.
 At this stage we use the same recipe as in  the parameterization of the qutrit selfcomplementary
 channels. One can parameterize the Kraus representation by introducing a phase $\theta$ and a set of
 parameters given by a unitary matrix $W$ of order $N$.
Let us introduce a rescaled unitary matrix $X=sW$ with $s\leq1$ such that $XX^{\dagger}=s^2 \1_N$. The completeness relation $ \sum_{i=1}^{k} {K'}_{i}^{\dagger}{K'}_{i} = {\bf 1}$ allows us to reduce the number of parameters. Finally, the structure of the Kraus operators reads
\begin{equation}
\begin{cases}
  K_1'=\begin{bmatrix} \cos\theta&0&\ldots&0\\0&\frac{1}{\sqrt{2}}\cos\theta&\ldots&0\\\vdots&\vdots&\ddots&\vdots\\0&0&\ldots&\frac{1}{\sqrt{2}}\cos\theta\end{bmatrix},\\
  K_i'=\begin{bmatrix} P^{i}[\frac{1}{\sqrt{2}}\cos\theta\ \ \ldots \ \ 0]\\(\frac{1}{\sqrt{N-2}}col(P^{-i}W,1)\sin\theta)^T\\\vdots\\(\frac{1}{\sqrt{N-2}}col(P^{-i}W,N-1)\sin\theta)^T\\(\frac{1}{\sqrt{N-1}}col(P^{-i}W,N)\sin\theta)^T\end{bmatrix}.\\
\end{cases}
\end{equation}
where $P$ denotes  the cyclic permutation matrix, 
$col(A,i)$ denotes the $i$-th column of the matrix $A$ and $T$ is the transposition.

Negativity (\ref{NegativitySys}) for presented generalized family of selfcomplementary maps is maximal for $\theta=0$. The Kraus operators in this case read   
\begin{equation}
\begin{cases}
  K_1'=diag\begin{bmatrix} 1&\frac{1}{\sqrt{2}}&\ldots&\frac{1}{\sqrt{2}}\end{bmatrix},\\
  K_i'=\begin{bmatrix} 0&\ldots&\frac{1}{\sqrt{2}}_{(i)}&\ldots&0\end{bmatrix},\\
\end{cases}
\end{equation}
where $(i)$ designates consecutive Kraus operators ($K_i \in M_{N,1}$) as well as $(i)$-th place in row where $1/\sqrt{2}$ is placed.

\end{document}